0bar

# An Overview of Video Allocation Algorithms for Flash-based SSD Storage Systems

Jaafer Al-Sabateen [1], Saleh Ali Alomari [2] and Putra Sumari [3]

[1,2,3] Multimedia Computing Research Group, School of Computer Science
University Sains Malaysia
11800 Pulau Pinang, Malaysia

*Abstract*—Despite the fact that Solid State Disk (SSD) data storage media had offered a revolutionary property storages community, but the unavailability of a comprehensive allocation strategy in SSDs storage media, leads to consuming the available space, random writing processes, time-consuming reading processes, and system resources consumption. In order to overcome these challenges, an efficient allocation algorithm is a desirable option. In this paper, we had executed an intensive investigation on the SSD-based allocation algorithms that had been proposed by the knowledge community. An explanatory comparison had been made between these algorithms. We reviewed these algorithms in order to building advanced knowledge armature that would help in inventing new allocation algorithms for this type of storage media.

*Keywords-SSDs; Allocation Algorithms; Data Management Systems; Garbage Collection; Storage Media.*

I. INTRODUCTION

The Solid State Disk (SSD) is a high performance data storage device that hasn't any moving parts, and achieved a superior performance comparing with traditional storage media [3, 6, 7, 16]. It's based on flash memory technology [7, 20]. Nowadays, flash-based Solid State Drive (SSD) has been widely used as a storage device for many systems such as laptops computers, enterprise servers and other digital devices [3, 5, 11]. The widely used were caused by many features that are available in this type of storage devices. For example, these systems have low-power consumption features, non-volatility properties, high random access performance, and high mobility platforms [8, 11, 19, 22]. For several years, there has been a significant growth in the flash-based Solid State Drive (SSD) market, due to the previous mentioned features [15, 17]. Recently, due to the sequential price reduction of flash-based storage systems, the Solid State Drive (SSD) is emerging as a killer application NAND flash in general purpose computing areas such as, desktop personal computers [11, 17], and enterprise servers [1, 11]. Technically, there are two basic types of SSD storage systems, Multi-Level Cell (MLC), and Single Level Cell (SLC) [11, 17, 21]. The main differences between these two types are concerning with the number of writing cycles [8, 10], and the storage capacity level, which mainly affected the SSD media life span [10, 14, 21, 22]. Physically, there are three main components that SSD data storage system consists of, the flash package, SSD controller and host interface logic [7, 12, 14]. These components are simplified and illustrated in the following figure.

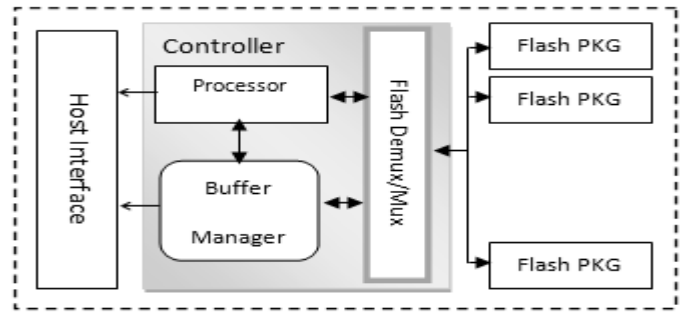

Figure 1. Basic Architecture of Solid State Disk (SSD) Storage Media

The controller as shown in figure 1, consist of three main parts, processor, buffer manager, and flash demux/mux integrated circuit [14]. The main function of the processor is to manage the flow of data and mappings from the logical block address to physical locations [5, 11, 14]. Respectively, the buffer manager will speed up the processing time required for performing several storage system functions, such as reading or writing operations [9, 10], while flash demux/mux is to managing instructions and data transport processes along the serial connections to the flash packages [13,15]. The host interface represents the point that connects internal environment for SSD storage system with the external physical environment. Regarding to package component, there are three basic terms should be clarified, Cell, Page, and Block components. The cell is the basic smallest unit of the block. The page consists of fixed number of cells, and a several sets of pages could be grouped into one block. An overall simplified architecture is shown in figure 2.

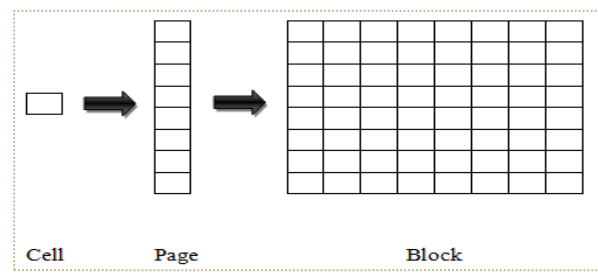

Figure 2. Block Basic Architecture

In this investigation paper, we will present a group of allocation algorithms that were proposed to handle some of challenges that emerged from three main Management issues in SSD storage systems, which are garbage collection process, limited erasure cycles, and out-place updating method [2, 4, 8 and 16].

The remainder of this overview study is organized into five main sections. Section (2) is the related background of flash-based SSD storage disk, which presents the storage system and explains the basic components that are forming it, clarifying some operational characteristics that executed in it such as out place updating scheme, garbage cleaning process, and limited erasure cycles is included there also. Section (3) shows the motivation factors that motivate the researchers to design and implement a set of flash-based allocation algorithms and the core problem that this investigation paper focuses on. The section (4) provides a literature review for allocation algorithms proposed for flash-based SSD system and suggested in the knowledge community. Regarding to section (5), it presents a comprehensive comparison between these algorithms and shows the main Discriminated Degree between it, and finally, section (6) clarify the conclusions for this investigation paper and general recommendations.

## II. RELATED BACKGROUND

The flash-based Solid State Disk (SSD) consists of a big number of blocks [17]. The read and write of data is done on cell basis while the erase is carried on block unit [1, 3, 19, 20, 22]. There are three basic operational characteristics, out place updating scheme, garbage cleaning process, and limited erasure cycles [3, 7, 8, 11, 16, 20, 21 ]. In flash-based Solid State Disk storage media, updating the existing data by overwriting the same physical location is strictly prohibited [11, 20, and 21]. The original data must be erased prior to the updated data can be stored on same location. To avoid from initiating the erase operation each time the data is updated, out place updating scheme has been recognized in flash-based storage media [3, 7, 8, 11, 16, 2 ]. In this scheme, the update data is written in new location while the old version is marked as invalid [1, 3, 7, 8, 11, 16]. After a long series of updating transaction occurs in the media, big volume of available space in flash-based storage media is consumed [6, 17, 22]. The garbage cleaning process is invoked when the ratio of invalid (garbage) reaches a certain level of threshold [2, 17]. Before the process can be initiated, valid data resided in the block must be copied into available free spaces of other blocks [1, 5, 21]. In this situation, additional processing time is consumed by garbage cleaning (collection) process [4, 5, 13, 19]. Moreover, system resources are consumed in form of system performance weakness. The erase operation is necessary in flash-based memory systems, in order to ensure the continuity of data storing process. Limited erasure cycles refer to number of erase operation allowed to each block [6, 11]. Each block has its own limitation in erasure access lifespan [7, 9, 19, 21, 22]. Excessively accessing will cause the block become unreliable and spoiled [9, 12, 15]. These factors affected on the performance of this media [3, 11, 22]. In The midst of this processes in the storage media, the need for allocation algorithm had become a critical option.

## III. ALLOCATION ALGORITHMS MOTIVATIONS

The flash-based Solid State Disk (SSD) storage media have three hardware characteristics, garbage collection mode; out-place updating scheme and limited erasure cycles, which affected its general performance [3, 7, 8, 11, 16, 20, 21]. The space waste, time consumption, and Random writing operations, had been appeared as a group of performance challenges that emerged from these characteristics. An allocation algorithm that mitigates its ruggedness should be invented and presented.

## IV. SSD-BASED ALLOCATION ALGORITHMS

To eliminate the ruggedness of the out place updating scheme, garbage cleaning process, and limited erasure cycles. Many of flash-based allocation algorithms had been proposed in the literature. Table 1 show these algorithms which had been proposed by different authors, this contribution came as a group of relief options that should mitigate the ruggedness degree of the SSD operational characteristics.

TABLE I. RESEARCH CONTRIBUTION FOR SSD-BASED ALLOCATION ALGORITHMS

| Algorithms | Publication | Author(s) |
|---|---|---|
| FCFS | 2005 | Li-Fu Chou, Pangfeng Liu[3] |
| FRFS | | |
| On-line FRFS | | |
| Probability-based | 2009 | Putra Sumari, Amir Rizaan Rahman [22] |
| Best-M | 2007 | Pangfeng Liu, Chung-Hao Chuang, and Jan-Jan Wu [19] |

However, next sections reviews and discusses these proposed allocations algorithms in more details.

### A. First Come First Serve (FCFS)

First Come First Serve (FCFS) [3] is a flash-based memory systems allocation algorithm. FCFS suggests the arrival time property for each incoming data page to be as base of allocation decision. FCFS allocates the modified data pages without any computation complexity. It uses a blocks list data structure which consider as memory blocks store. For the incoming modified data pages, FCFS algorithm selects the appropriate block from this list. In general allocation procedure of FCFS algorithm, the algorithm places the first modified data page into the first position of the first block of the selected block that had selected from block list data structure. The second page is placed into the second position of the first blocks of the selected block from block list data structure, and so on. FCFS allocation algorithm, places the data pages into blocks in sequentially style. The following Pseudo code explains how allocation decision is made during this algorithm:

> 1. *for each page in the page access pattern*
> 2. *Place the page into the first available cell from the block list.*
> 3. *If the page appeared before then:*
> 4. *mark the cell it previously resided INVALID*
> 5. *If the block the page previously resided now becomes INACTIVE then:*
> 6. *erase it and move it to the end of the block list*

Algorithm 1. Pseudo Code for First Come First Serve (FCFS) Allocation Algorithm

One of the most important steps in allocation decision is to set the position of the page according to page arrival time. The allocation (placing) decision in this algorithm is based on the equation below:

***(POSITION – B\*R)%B)* [3], where B is a Number of Pages in the Block, and R is a Number of the Erased Bocks.**

Page arrival time refer to the time of the page when had been appeared in page access pattern. FCFS, initializing the block list then it set the position for the incoming pages according to their arrival time. Because FCFS places pages according to their arrival time, a data page with lowest arrival time value will be placed at first, and then the page with greater arrival time value will be the next page and so on. However, if the page appears another time in the same block, then FCFS algorithm will mark the old page as invalid one, the system in this case recognize the last page as the valid one [3].

*B. First Re-Arrival First Serve (FRFS)*

First Re-arrival First Serve (FRFS) [3] allocation algorithm had been emerged from previous FCFS allocation algorithm. Instead of placing data pages according to their arrival time property, FRFS algorithm depends on calculating the re-arrive time value for each incoming file data page in order to determine the correct position (location) for it. Re arrival time is the time when the page re-arrives. FRFS allocation algorithm, uses the minimal number of blocks, due to it reuses block as soon as possible. FRFS algorithm has three main allocation stages, in first stage, it computes the re-arrival time of each page by scanning through the entire access pattern and then in the second stage, it allocates a cell for each data page. The page having the earliest re-arrival time is assigned (0) value, the page with the second earliest re-arrival time is assigned (1) value, and so on. In the third stage it places the pages into the blocks according to the ordinal number they are assigned from the second stage. If this algorithm determines that a block had become in inactive state, it will erase the block and moving it to the end of the block list data structure so it can be reused later. FRFS keeps track of number of active blocks, and at the end, the maximum of these active block numbers is the number of blocks required by it [3]. The following presents the pseudo code for First Re-arrival First Serve allocation (FRFS) algorithm:

1. *Initialize the block list*
2. *Compute the re-arrival time for each block*
3. *Compute the order according to the re-arrival time*
4. *For each page in the page access pattern*
5. *Set the position of the page according to the order of the page*
6. *Place the page into the ((position-B\*R)%B)$^{th}$ cell of the ((position-R\*B)/B)$^{th}$ block in the block list*
7. *If the page appeared before*
8. *mark the cell it resided invalid*
9. *If the block where the page resided before is now inactive*
10. *move it to the end of the block list to be reused*
11. *count the number of the active block in the block list*
12. *compare it with the one from the previous iteration*

Algorithm 2. The Pseudo Code for First Re-arrival First Serve (FRFS) Allocation

Regarding to allocation procedure, First Re-arrival First Serve (FRFS) algorithm start by initializing the block list, and then it executes calculation process to compute the re-arrival time for each page. The placing pages procedure is based on the following equation:

***(POSITION – B\*R)%B)* [3], where B: Number of Pages in the Block, R: Number of the Erased Bocks.**

*C. Online First Re-Arrival First Serve (On-Line Frfs)*

Online First Re-arrival First Serve [3] allocation algorithm should analyze the entire page access pattern in order to make the allocation decisions. It takes the allocation decision as soon as the page request arrives. Two essential data structure are used in this algorithm, block list data structure, and the second one is a prediction table that contains prediction information for all the pages that have appeared. The block list data structure contains all the blocks of the memory and the blocks there are sorted by their identification numbers [3]. The prediction table data structure contains information for all pages that have been appeared. To access the prediction information faster, the prediction table should be placed in high speed processing time memory such as Random Access Memory (RAM). According to what previously mentioned, each page has two data type, the estimated arrival interval for the page and last time that the page appeared. When new data page appears, On-Line FRFS algorithm estimates arrival interval and set it to the default value, the default value, in this case, refers to the mean value of the intervals of all pages that had been observed in the past, then On-Line FRFS the last arrival time of the page to current time. After that, On-Line FRFS insert this entry into the prediction table. Then On-Line FRFS will update the estimated interval value and the length of interval between the current time and the previously arrival time of the page [10].

Mathematically, the (On-Line FRFS) algorithm computes the interval estimated length using the following equation:

*New estimated arrival interval = **R\*ER + (1-R)\*(T-LT)** [3], where: R is a Constant between (0) and (1), ER is the Old Estimated Interval of the Data Page, T is the Current Time, and LT is the Data Page Last Arrival Time.*

Many mathematical steps should be followed in order to determine the correct position for incoming data pages, the following pseudo code shows the basic allocation steps that data pages should be passed through during allocation process:

1. *compute the re-arrival time for each page*
2. *compute an order of re-arrival time*
3. *for each page in page access pattern*
4. *let K be the ordinal number*
5. *place the page into the K/B$^{th}$ block in the block list*
6. *if the same page appeared before*
7. *mark the cell it resided invalid*
8. *if the block the page previously resides became inactive*
9. *erase and move it to the end of the block list*

Algorithm 3. Pseudo Code of On-Line First Re-arrival First Serve Allocation Algorithm

*D. Probability Based Popularity Allocation Scheme*

Probability based popularity [22] allocation algorithm consider as one of modern allocation algorithm. The allocation decision is based on the probability of the page which basically depend on the popularity of each unique page in the pages access pattern. Regarding to this algorithm, the distribution of pages in the page access pattern is assumed to follows the (ZIPFs Law) [22], where pages are stored in an increasing order, from the heights rank to the lowest rank. This distribution is based on pages frequency of the occurrences in the access pattern [22].

Depending on the number of times that each page is appearing into page access pattern, this algorithm divides the blocks into two categories HOT and COLD blocks. Probability based popularity allocation algorithm consist of two main components, access screening and allocation algorithm. Access screening percolate the type of data access to two categorization, read and write, the allocation algorithm part perform a distribution process of the access data. This component consist of two sub component, popularity interpreter and allocation engine.

Two main functions are performed by these sub components. Popularity interpreter determines the state of accessed data, either hot or cold. The allocation engine performs block selection process [22]. The distribution process performed by the allocation algorithm component will cluster the distribution into three main groups, maximum probability, median probability and minimum probability [22].

In this state, the pages that have probability value greater than or equal to median probability are classified as hot page, others page will be classified to cold pages. Probability Based allocation algorithm, divide the blocks of flash-based storage media into two categories, hot and cold blocks. The probability based popularity algorithm assumes that the number of hot blocks should be more that number of cold blocks. The amount of blocks which are subjected to be hot or cold is calculated using the following two equations:

*1. H = (MAX+MEDIAN)\*B.*
*2. C = 1- H [22].*

Where, H is the Amount of HOT blocks, MAX is the maximum probability, MEDIAN is the median probability, B is the cod block position and C is the Amount of COLD blocks.

Regarding to above mentioned equations, The Probability based popularity allocation algorithm keeps each block not to be in active state in the whole allocation process. If such blocks are active during the whole allocation process then number of active blocks will be increased. The overall allocation algorithm is simplified as follow:

1. *get the probability of each page*
2. *set the highest probability to MAX*
3. *set the lowest probability to MIN*
4. *get the middle probability and set the MEDIAN*
5. *for each page in the access pattern*
6. *get the page weight*
7. *if weight >= MEDIAN*
8. *set page as HOT*
9. *write page into free pages within blocks [0,H 1]*
10. *else*
11. *Set page as COLD*
12. *Write page into free pages within block [H,B 1]*
13. *count ACTIVE blocks*

Algorithm 4. Pseudo Code for Probability Based Allocation Algorithm

*E. Best-M Algorithm*

Best-M allocation algorithm is stand on Block-Based Allocation concept. Simply it places all the requests for the same pages into the same block. As a result, each block contains only a single most up-to-date content of a page plus all the previous contents that all have been marked invalid. The idea of this algorithm is to assign page appearances to cells according to their difference. A difference between two page appearances is defined as the sum of the difference of their arrival time and the difference of their re-arrival time. The reason that it uses difference to allocate cells is that it is likely that all cells in the same block will be set to valid and invalid at about at the same time [19]. Best-M Algorithm could be simplified by the following pseudo code:

1. *compute block index for page appearance*
2. *initialize block list*
3. *search block list for block(i)*
4. *If block(i) not found in the block list*
5. *insert the new block in the block list*
6. *set index(i) for this block*
7. *Place page into first cell of block(i)*
8. *If block became inactive*
9. *delete block index from block list(i)*

Algorithm 5. The Pseudo Code for Best-M Algorithm.

Best-M allocation algorithm uses block list data structure to store blocks in it. The block in Best-M allocation algorithm has two properties, block index and number of cells for each block. When page arrive, Best-M algorithm, compute block index (i) for this page, then it search block list to find this block. If the requested block not found in the block list, Best-M insert new block in the block list and assign index (i) to it. After that, Best-M places the page into that block. However, block recycle process is invoked when the state of block became Inactive [11].

## V. COMPARISION BETWEEN FLASH-BASED SSD ALLOCATION ALGORITHMS

A group of algorithms had been discussed and various allocation techniques were adopted by each previous mentioned algorithm. Table1 show a general review for these algorithms then we will compare these algorithms theoretically in order to clarify and explain the main differences between it.

TABLE II. A GENERAL COMPARISON BETWEEN ALLOCATION ALGORITHMS USED IN SSD STORAGE SYSTEMS.

| Comparison factor | Algorithm | | | | |
|---|---|---|---|---|---|
| | *FCFS* | *FRFS* | *Online FCFS* | *Probability base* | *Best_M* |
| **Algorithm Type** | On Line | Off Line | On Line | On Line | Off Line |
| **Allocation mode** | Arrival time | Re-arrival time | Pages Arrival Interval | Page probability | Difference value |
| **Futures Discriminated Degree** | Sequential allocation style | Emerged from FCFS | Mathematical based allocation process | Divide blocks into HOT & COLD | Each block with same data pages |

As shown in table 2, there are five main algorithms. Online and Offline are two main types of allocation scenarios. The Online type referring to online allocation mode and Offline type is the internal-style data real architecture style. The allocation mode factor explains the concept that the allocation procedure is based on. Generally, future discriminated degree referring to the basic concept(s) that were adopted to perform the allocation procedure. We had executed these algorithms in real environment (simulation environment). The results presented in table 3, showing the complexity of each algorithm. The complexity consists of two main parameters, space and execution time. The space criterion is referring to the number of cells that had been requested by the storage system when (n) number of incoming pages had to be written. The execution time referring to the time required to execute the specified (pre-defined) number of write operations that are waiting in the page access pattern queue.

TABLE III. EMPIRICAL COMPARISON BETWEEN ALLOCATION ALGORITHMS USED IN SSD STORAGE SYSTEMS.

| Algorithm | Complexity | | Proposed work |
|---|---|---|---|
| | *Space required* | *Execution time* | |
| **First Come First Serve** | O (nn) | O (mn) | allocation algorithm using arrival time parameter |
| **First Re-arrival First Serve** | O (nn) | O (mn) | allocation algorithm using re-arrival time parameter |
| **On-line FRFS** | O (n) | O (n3) | On-line allocation algorithm |
| **Probability-based** | O (mn) | O (n2) | allocation algorithm based on page popularity probability |
| **Best Match (Best-M)** | O (n) | O (n2) | block-based allocation algorithm using page appearance and cell difference to parameters |

As mentioned in table 3, the probability-based, and Best Match (Best-M) allocation algorithms, had satisfied best degree comparing with other allocation algorithms, despite that fact that First Come First Serve, and First Re-arrival First Serve (FRFS), had satisfied less execution time comparing with other algorithms.

## VI. CONCLUSION

The paper provided an overview of the available literature in data allocation techniques for flash-based SSD storage systems. We have introduced, analyzed and compared several allocation techniques. We highlighted the importance of allocation schemes in both reducing waste-space, and adjust the random-based style of data writing operations. Respectively, the findings were objectively reported and it may considered as researching reference for obtaining obviously documented knowledge, on many flash-based allocation strategies. Generally, as a conclusion, the storage devices with built-in allocation algorithm, performs better than those haven't one. More deeply future work on this type of allocation strategies is required.

ACKNOWLEDGMENT

This work was supported by a grant from the Universiti Sains Malaysia. We thanks and recognition go to my advisor, Associate Professor. Dr. Putra Sumari, who for helping us in this paper. Last but not least, the authors would like to thank the School of Computer Science, Universiti Sains Malaysia (USM) for supporting this study.

AUTHORS PROFILE

**Jaafer Mohammad Al-Sabateen** is a researcher in University Science Malaysia, School of Computer Science. He obtained software engineering Bachelor degree from Isra private university in 2008 and continued his study to a master degree in Computer Science.

**Saleh Ali Alomari** obtained his Bachelor degree in Computer Science from Jerash University, Jordan in 2005 and Master degree in Computer Science from Universiti Sains Malaysia (USM), Pulau Penang, Malaysia in 2007. Currently, He is a Ph. D. candidate at the School of Computer Science, Universiti Sains Malaysia. He is the candidate of the Multimedia Computing Research Group, School of Computer Science, USM. He is managing director of ICT Technology and Research and Development Division (R&D) in D&D Professional Consulting Company. He has published over 35 papers in international journals and refereed conferences. He is a member and reviewer of several international journals and conferences (IEICE, ACM, KSII, JDCTA, IEEE, IACSIT, etc). His research interest are in area of Multimedia Networking, video communications system design, multimedia communication specifically on Video on Demand system, P2P Media Streaming, MANETs, caching techniques and for advanced mobile broadcasting networks as well.

**Putra Sumari** obtained his MSc and PhD in 1997 and 2000 from Liverpool University, England. Currently, he is a lecturer at the School of Computer Science, Universiti Sains Malaysia, Penang. He is the head of the Multimedia Computing Research Group, School of Computer Science, USM. Member of ACM and IEEE, Program Committee and reviewer of several International Conference on Information and Communication Technology (ICT), Committee of Malaysian ISO Standard Working Group on Software Engineering Practice, Chairman of Industrial Training Program School of Computer Science USM, Advisor of Master in Multimedia Education Program, UPSI, Perak.